\documentstyle[12pt]{article}
\def\beq{\begin{equation}}
\def\eeq{\end{equation}}

\def\beqn{\begin{eqnarray}}
\def\eeqn{\end{eqnarray}}
\relax

\relax
\catcode`\@=11
\@addtoreset{equation}{section}

\def\@normalsize{\@setsize\normalsize{15pt}\xiipt\@xiipt
\abovedisplayskip 14pt plus3pt minus3pt%
\belowdisplayskip \abovedisplayskip
\abovedisplayshortskip  \z@ plus3pt%
\belowdisplayshortskip  7pt plus3.5pt minus0pt}
\def\small{\@setsize\small{13.6pt}\xipt\@xipt
\abovedisplayskip 13pt plus3pt minus3pt%
\belowdisplayskip \abovedisplayskip
\abovedisplayshortskip  \z@ plus3pt%
\belowdisplayshortskip  7pt plus3.5pt minus0pt

\def\@listi{\parsep 4.5pt plus 2pt minus 1pt
            \itemsep \parsep
            \topsep 9pt plus 3pt minus 3pt}}

\def\underline#1{\relax\ifmmode\@@underline#1\else
	$\@@underline{\hbox{#1}}$\relax\fi}
\@twosidetrue

\catcode`@=12

\catcode`\@=11

\catcode`\@=12

\relax

\def\figcap{\section*{Figure Captions\markboth
	{FIGURECAPTIONS}{FIGURECAPTIONS}}\list
	{Fig. \arabic{enumi}:\hfill}{\settowidth\labelwidth{Fig. 999:}
	\leftmargin\labelwidth
	\advance\leftmargin\labelsep\usecounter{enumi}}}
 \relax
\def\FERMIPUB{}
\def\FERMILABPub#1{\def\FERMIPUB{#1}}
\def\ps@headings{\def\@oddfoot{}\def\@evenfoot{}
\def\@oddhead{\hbox{}\hfill
	\makebox[.5\textwidth]{\raggedright\ignorespaces --\thepage{}--
	\hfill {\rm FERMILAB--Pub--\FERMIPUB}}}
\def\@evenhead{\@oddhead}
\def\subsectionmark##1{\markboth{##1}{}}
}

\ps@headings
\relax
\def\sp#1{{\rm Tr}\biggl( #1 \biggr)}
\def\Ud{U^\dagger}
\def\ra{\rightarrow}
\FERMILABPub{93/024--T}
\begin{document}
\begin{titlepage}
\def\ba{\begin{array}}
\def\ea{\end{array}}
\def\thefootnote{\fnsymbol{footnote}}
\begin{flushright}
	FERMILAB--PUB--93/024--T\\
	February 1993
\end{flushright}
\vfill
\begin{center}
{\large \bf $K_L \ra \mu^\pm e^\mp \nu \overline{\nu}$
as background to $K_L \ra \mu^\pm e^\mp$}\\
\vfill
	{\bf S.~Dawson$^{(a)}$}\footnote{
This work was supported by the U.S. Department of Energy under contract
DE-AC02-76CH00016.}{\bf  and G.~Valencia$^{(b)}$}\\
{\it  $^{(a)}$ Physics Department\\
               Brookhaven National Laboratory\\ Upton, NY 11973}\\
{\it  $^{(b)}$ Theoretical Physics\\
               Fermi National Accelerator Laboratory \\
               Batavia, IL 60510}\\
\vfill
\end{center}
\begin{abstract}

We consider the process $K_L \ra \mu^\pm e^\mp \nu \overline{\nu}$ at next to
leading order in chiral perturbation theory. This process occurs in the
standard model at second order in the weak interaction and constitutes a
potential background in searches for new physics through the modes
$K_L \ra \mu^\pm e^\mp$. We find that the same cut, $M_{\mu e}>489$~MeV,
used to remove the sequential decays $K_{l3}\ra \pi_{l2}$ pushes the
$B(K_L \ra \mu^\pm e^\mp \nu \overline{\nu})$ to the $10^{-23}$ level,
effectively removing it as a background.

\end{abstract}

\end{titlepage}

\clearpage

\section{Introduction}

In the minimal standard model with massless neutrinos, the lepton
family number is absolutely conserved so the decays $K_L \ra \mu^\pm e^\mp$
do not occur. Their observation would constitute clear evidence for
new physics. In view of this, there have been several experiments
studying these modes and others are planned for the future. The current
experimental upper bounds are:
\beq
B(K_L \ra \mu^\pm e^\mp) < \cases{
9.7 \times 10^{-11} & KEK-137 \cite{mue137} \cr
3.3 \times 10^{-11} & AGS-791 \cite{mue791} \quad .\cr}
\label{elimit}
\eeq
This latter number represents the most sensitive kaon experiment to date.
The proposed successor experiment, AGS-871, expects to improve the
sensitivity by a factor of about 20.

A model independent study of this type of processes can be done following the
approach of Buchm\"{u}ller and Wyler, Ref.~\cite{buwy}.
The physics beyond the standard
model is parameterized by an effective Lagrangian that is gauge invariant
under $SU_c(3)\times SU_L(2)\times U_Y(1)$.
This Lagrangian is given by a sum of four
fermion operators, of which we present the purely left handed one as
as example:
\beq
{\cal O}_{V-A}= C_{V-A} {g^2 \over  \Lambda^2}\overline{\mu}
\gamma_\mu {(1+\gamma_5 )\over 2} e \overline{s}
\gamma^\mu {(1+\gamma_5 )\over 2} d \quad .
\label{efop}
\eeq
The factor $g^2$ is included to reflect the fact that we think of these
operators as originating in the exchange of a heavy
gauge boson (or perhaps a scalar) in the new physics sector.
Any additional factors, like mixing angles, are contained in
the coefficient $C_i$. It is conventional to assume that
$C_i$ is of order ${\cal O}(1)$, so that $\Lambda$ is the scale
that characterizes the heavy degrees of freedom
(typically the mass of the exchanged boson).
One then interprets the bounds on the decays
induced by these operators as bounds on the ``scale of new physics''
$\Lambda$.

It is standard to compare this mode to the rate for $K^+ \ra \mu^+ \nu$,
to absorb the hadronic matrix element as well as common kinematical factors
in the limit $m_e =0$. One finds:
\beq
{\Gamma (K_L \ra \mu^+ e^-) \over \Gamma(K^+ \ra \mu^+ \nu)} =
2{C^2_{V-A}\over |V_{us}|^2} \biggl({m_W \over \Lambda}\biggr)^4
\label{rmeva}
\eeq
The current experimental limit Eq.~\ref{elimit} then implies
the bound $\Lambda > 108~TeV$. This result can be interpreted as
a bound on the mass of new particles in different models \cite{beyondsm}.

These decays are also allowed in minimal extensions of the Standard Model
in which the neutrinos are given a mass. The decays then proceed via
one-loop box diagrams. The decay rate for $K_L \ra \mu^\pm
e^\mp$ is proportional to the product of mixing angles between the $\mu$, $e$
and the heavy neutral lepton $N$: $|U_{Ne}U^*_{N\mu}|^2$. Using the
result $B(\mu \ra e \gamma)< 4.9 \times 10^{-11}$ \cite{mueexp}, the authors
of Ref.~\cite{acpak} find $|U_{Ne}U^*_{N\mu}|^2 < 7 \times 10^{-6}$ for
$m_N > 45~GeV$. From this they conclude that
$B(K_L \ra \mu^\pm e^\mp)$ is at most
$10^{-15}$ in this type of models. Marciano \cite{marhn} has pointed
out that there is a better bound  $|U_{Ne}U^*_{N\mu}|^2 < 10^{-8}$
coming from $ \mu  \ra e $ conversion in the field of a heavy nucleus.
With this bound one finds $B(K_L \ra \mu^\pm e^\mp)$ to be at most
a few times $10^{-18}$. In left-right symmetric models
$B(K_L \ra \mu^\pm e^\mp)$ can be as
large as $10^{-13}$, although this happens only
in a small corner of parameter space \cite{babrbe}.

Given the level of sensitivity expected for future experiments, as well
as the small rates predicted in many models, it is important to study
processes that could fake this one at the $10^{-15}$ level.
The ultimate background for the decays $K_L \ra \mu^\pm e^\mp$
is the standard model process $K_L \ra \mu^\pm e^\mp \nu_\mu \nu_e$,
which we study in this paper using the techniques of chiral
perturbation theory ($\chi$PT).

\section{Chiral Lagrangian}

The lowest order chiral Lagrangian, ${\cal O} (p^2)$, is \cite{cpta,cptb}:
\beq
{\cal L}^{(2)}_S = {f^2_\pi \over 4}\sp{D_\mu U D^\mu \Ud}
+ B_0 {f^2_\pi \over 2}\sp{MU+\Ud M}.
\label{slt}
\eeq
$M$ is the diagonal matrix $(m_u,m_d,m_s)$, and
the meson fields are contained in the matrix $U=\exp(2i\phi /f_\pi)$
with:
\beqn
\phi &=& {1 \over \sqrt{2}}
\left( \begin{array}{ccc}
\pi^0 /\sqrt{2}+\eta /\sqrt{6} & \pi^+ & K^+ \\
\pi^- &-\pi^0 /\sqrt{2}+\eta /\sqrt{6} & K^0 \\
K^- & \overline{K^0} & 2 \eta /\sqrt{6}
       \end{array} \right) \quad .
\label{pions}
\eeqn
$U$ transforms under the chiral group as $U \ra R U L^\dagger$.
For our purpose, the charged $W^\pm$ bosons will be external
fields, contained in the covariant derivative:
\beqn
D_\mu U &=& \partial_\mu  U + i U l_\mu \nonumber \\
l_\mu &=& -{e \over \sqrt{2} \sin\theta_W}
\biggl(W^+_\mu T + W^-_\mu T^\dagger \biggr),
\label{covd}
\eeqn
where $T$ is the matrix:
\beq
T=
\left( \begin{array}{ccc}
0     &     V_{ud}     &   V_{us}  \\
0   &  0    &   0   \\
0   &  0    &   0
       \end{array} \right).
\label{gencd}
\eeq
For the pion decay constant we use $f_\pi =93$~MeV.

At next to leading order, ${\cal O}(p^4)$, there are ten more operators
\cite{cptb} in the normal intrinsic parity sector.
For the process we consider only one out of the ten terms contributes:
\beq
{\cal L}^{(4)}_S = -iL_9 \sp{F_L^{\mu \nu}D_\mu \Ud D_\nu U}
\label{slf}
\eeq
Since there are no photons involved, the field strength tensor $F_{\mu\nu}$
is given by:
\beq
F^{\mu\nu}_L = \partial_\mu l_\nu - \partial_\nu l_\mu - i
[l_\mu,l_\nu] \quad .
\eeq
 At this same order there
is also a contribution from the anomaly.
The contribution from the anomaly is given by
the Wess-Zumino-Witten anomalous action \cite{wzw}. It contributes the
following term to the process we study:
\beq
{\cal L}^{(4)}_{WZW} = -{g^2 N_c \over 48 \pi^2 f_\pi}{\rm Re}\left(
V_{ud} V^*_{us} \right) \epsilon^{\mu\nu\alpha\beta}
\biggl(\partial_\mu W^-_\nu W^+ _\alpha +
W^-_\mu \partial_\nu W^+_\alpha \biggr) \partial_\beta K_2^0,
\label{wzwl}
\eeq
where $K^0_2$ is the CP odd neutral kaon (we ignore CP violation)
and $N_c=3$.

A complete  calculation to ${\cal O}(p^4)$ consists of tree-level diagrams
with vertices from Eqs.~\ref{slt}, \ref{slf}, and \ref{wzwl}, and
of one-loop diagrams using only Eq.~\ref{slt}. For the purpose of our
paper it will be sufficient to ignore the one-loop contributions. In order
for this to be consistent we use values for $L_9$ that are scale independent
and derived from tree-level models. In particular, we will use
$L_9 = 7.3 \times 10^{-3}$ from vector meson resonance saturation models
and $L_9 = 6.3 \times 10^{-3}$ from quark models.

\section{$K_L \ra \mu^\pm e^\mp \nu \overline{\nu}$}

To leading order in $\chi$PT this process is dominated by
$K_L \ra \pi^\pm e^\mp \nu_e$ followed by $\pi^\pm \ra \mu^\pm \nu_\mu$. The
branching ratio for this chain can be estimated using the narrow
width approximation to be a huge $38\%$ (we have summed over the two
modes). The processes with
$\mu \leftrightarrow e$ exchanged are helicity suppressed and are
typically ignored. It is easy to see
that the maximum invariant mass of the lepton pair in this
sequential decay is $m_{\mu e}< 489~MeV$.
It is therefore possible to remove this background with a cut
on the lepton pair invariant mass. Going beyond the narrow width
approximation, and including next to leading order terms in $\chi$PT
can yield a lepton pair invariant mass larger than $489~MeV$.

In terms of the momenta of the particles involved,
$K_L(k) \ra e^-(p_e) \overline{\nu}_e(q_e) \mu^+ (p_m) \nu_m(q_m)$ it is
convenient to define $p^+ = q^m + p^m$, $p^- = q^e + p^e$, and to use
the variables:
\beq
y_m = {({p^+})^2 \over m^2_K}\;
 ,\;\; y_e = {({p^-})^2 \over m^2_K}\;,
\eeq
\beq
\nu = {2 k \cdot (q_e - p_e) \over m^2_K}\;.
\eeq
We normalize all masses to the $K_L$ mass defining for each particle
$P=e,\mu,\pi^+,K^+$, $r_P \equiv m_P /m_{K_L}$.
We also use the leptonic currents:
\beqn
L^e_\mu &=& \overline{u}_e \gamma_\mu (1 - \gamma_5)v_\nu \nonumber \\
L^m_\mu &=& \overline{u}_\nu \gamma_\mu (1 - \gamma_5)v_m \nonumber \\
\hat{L}^m &=& \overline{u}_\nu (1 + \gamma_5) v_m
\quad .
\eeqn
The lowest order matrix element is given by the sum of a charged pion and a
charged kaon poles as in Fig.~1a:
\beq
M^{(2)} = G_F^2 s_\theta c_\theta \biggl( {f_\pi \over y_m-r_\pi^2+
i \Gamma_\pi r_\pi} +{f_K \over y_m -r^2_{K^+}+i \Gamma_{K^+} r_{K^+}}\biggr)
{m_\mu \over m^2_K} (p^+\cdot L^e)\hat{L}^m
\eeq
where $s_\theta$ is the sine of the Cabibbo angle, $s_\theta \approx .22$.
The explicit factor $m_\mu$ reflects the helicity suppression that allows us
to drop the diagrams with $\mu -e$ interchanged.
Neglecting the electron mass, we then find:
\beq
\Gamma = {G_F^4 m^2_\mu m^5_{K_L}f^2_\pi \over 1536 \pi^5}
|s_\theta c_\theta|^2 I
\eeq
with
\beqn
I &=& \int_{r^2_\mu}^1 {dy \over y} (y-r^2_\mu)^2\biggl[-3y^2\log y +
{1\over 4}(1-y^2)(1-8y+y^2)\biggr]\nonumber \\
&& \cdot
\biggl| {1 \over y-r^2_\pi+ir_\pi {\Gamma_{\pi^+} \over m_{K_L}}}
             + {(f_K / f_\pi) \over y-1+ir_{K^+} {\Gamma_{K^+} \over m_{K_L}}}
\biggr|^2
\eeqn
The last integral can be performed analytically if we use the narrow
width approximation. Doing so we find a
$B(K_L\ra \mu^+ e^- \nu \overline{\nu})$ of $18\%$ from the $\pi^+$ pole,
and $1.8 \times 10^{-9}$ from the $K^+$ pole. However, noted before,
the largest $\mu-e$ invariant mass from these contributions is
$M^2_{\mu e}(max) = m^2_K + m^2_\mu - m^2_P$. Therefore $M_{\mu e} < 489$~MeV
for the pion pole and $M_{\mu e} < 123$~MeV for the $K^+$ pole.

We can estimate the size of the off-shell contributions from the lowest
order matrix element, by implementing a ``theorist's cut'' of $\pm 1$~MeV
around the pole mass. We then obtain
$B(K_L\ra \mu^+ e^- \nu \overline{\nu}) = 1.62 \times 10^{-15} $.
However, when we use the realistic cut $M_{\mu e} > 489$~MeV
instead, this number is dramatically
reduced to $B(K_L\ra \mu^+ e^- \nu \overline{\nu}) = 8.7 \times 10^{-24}$.

At next to leading order in $\chi$PT, we obtain the first structure
dependent contributions depicted schematically in Fig.~1b.
We find:
\beq
M^{(4)} =  {G^2_F \over f_\pi} s_\theta c_\theta \biggl\{ 4 L_9 \biggl[
m_K^2(y_e -y_m)(L^e \cdot L^m) - m_\mu (p^+ \cdot L^e)\hat{L}^m\biggr]
-{i \over 2\pi^2 } \epsilon^{\mu\beta
\rho\sigma}p^+_\mu p^-_\beta L^m_\rho L^e_\sigma \biggr\}
\label{seorme}
\eeq
where the first term comes from Eq.~\ref{slf}, and the second term
comes from the anomaly Eq.~\ref{wzwl}.

Beyond leading order, the rate receives contributions from Eq.~\ref{seorme}
and from its interference with the lowest order matrix element.
With no cuts, we find numerically (omitting the lowest order
$p^2$ contribution):
\beq
B^{(4)}(K_L\ra \mu^+ e^- \nu \overline{\nu}) =
5.1 \times 10^{-15} L_9 +
8.4 \times 10^{-14} L_9^2 +
4.5 \times 10^{-18}
\label{ntlore}
\eeq
With a cut $M_{\mu e}> 489$~MeV, and taking $L_9 = 0.007$ we find for
the complete rate to ${\cal O}(p^4)$
$B(K_L\ra \mu^+ e^- \nu \overline{\nu}) \approx 9 \times 10^{-24}$.

Although the standard model process $K_L\ra \mu^+ e^- \nu \overline{\nu}$
is a potential background for the lepton flavor number violating decay
$K_L \ra \mu^+ e^-$, we have found that the cut in the invariant mass of the
lepton pair $M_{\mu e} > 489$~MeV, necessary to remove the sequential
decays $K_{\ell 3} \ra \pi_{\ell 2}$ background, also
suppresses this decay well below expected sensitivities.

\noindent{\bf Acknowledgements}

We are grateful to L.~Littenberg and W.~Marciano for useful conversations.

\begin{figcap}

\item Diagrams contributing to $K_L \ra \mu^\pm e^\mp \nu \overline{\nu}$.
a) Pole diagrams at order ${\cal O}(p^2)$. b) Direct, or structure
dependent contributions at order ${\cal O}(p^4)$.

\end{figcap}

\end{document}